\documentclass[fleqn,11pt,twoside]{article}

\usepackage{amsmath,amsthm,amssymb,amscd,ifthen}

\makeatletter

\newcommand{\ArticleLabel}{Article label}
\newcommand{\evenhead}{Author \ name}
\newcommand{\oddhead}{Article \ name}
\newcommand{\theArticleName}{Article name}
\newcommand{\theAuthorNameForContents}{Author}
\newcommand{\theArticleNameForContents}{Article}

\newcommand{\FirstPageHeading}[2]{
\renewcommand{\ArticleLabel}{#1}
\newpage\thispagestyle{empty}\label{\ArticleLabel-fp}%
\noindent\raisebox{24pt}[0pt][0pt]{\makebox[\textwidth]{\protect\footnotesize \sf 
Proceedings of 4th Workshop ``Group Analysis of Differential Equations \& Integrability'' 
\hfill 2009
}}\par}

\newcommand{\LastPageEnding}{\label{\ArticleLabel-lp}\newpage}

\newcommand{\AuthorNameForContents}[1]{\renewcommand{\theAuthorNameForContents}{#1}}
\newcommand{\ArticleNameForContents}[1]{\renewcommand{\theArticleNameForContents}{#1}}
\newcommand{\ArticleName}[1]{\renewcommand{\theArticleName}{#1}\vspace{-2mm}\par\noindent {\LARGE\bf  #1\par}}
\newcommand{\Author}[1]{\vspace{5mm}\par\noindent {\it #1} \par\vspace{2mm}\par}
\newcommand{\Address}[1]{\vspace{2mm}\par\noindent {\it #1} \par}
\newcommand{\Email}[1]{\ifthenelse{\equal{#1}{}}{}{\par\noindent {\rm E-mail: }{\it  #1} \par}}
\newcommand{\EmailD}[1]{\ifthenelse{\equal{#1}{}}{}{\par\noindent {$\phantom{\dag}$~\rm E-mail: }{\it  #1} \par}}
\newcommand{\Abstract}[1]{\vspace{6mm}\par\noindent\hspace*{8mm}%
\parbox{120mm}{\small #1}\par\vspace{6mm}
\addtocontents{top}{\small\hangindent=10mm\hangafter=1\protect{\it \theAuthorNameForContents}, 
\protect{\rm \theArticleNameForContents}\dotfill\pageref{\ArticleLabel-fp}
\par\vspace{1mm}\par}\par
\addtocontents{tor}{\small\hangindent=10mm\hangafter=1\protect{\it \theAuthorNameForContents}, 
\protect{\rm \theArticleNameForContents}\dotfill\pageref{\ArticleLabel-fp}
\par\vspace{1mm}\par}\par
}
\newcommand{\tableofpapers}{\@starttoc{top}}
\newcommand{\tableofpapersA}{\@starttoc{tor}}

\newcommand{\ShortArticleName}[1]{\renewcommand{\oddhead}{#1}}
\newcommand{\AuthorNameForHeading}[1]{\renewcommand{\evenhead}{#1}}

\renewcommand{\@evenhead}{
\hspace*{-3pt}\raisebox{-15pt}[\headheight][0pt]{\vbox{\hbox to \textwidth 
{\thepage \hfil \evenhead}\vskip4pt \hrule}}}
\renewcommand{\@oddhead}{
\hspace*{-3pt}\raisebox{-15pt}[\headheight][0pt]{\vbox{\hbox to \textwidth 
{\oddhead \hfil \thepage}\vskip4pt\hrule}}}
\renewcommand{\@evenfoot}{}
\renewcommand{\@oddfoot}{}

\long\def\@makecaption#1#2{%
  \vskip\abovecaptionskip
  \sbox\@tempboxa{\small \textbf{#1.}\ \ #2}%
  \ifdim \wd\@tempboxa >\hsize
    {\small \textbf{#1.}\ \ #2}\par
  \else
    \global \@minipagefalse
    \hb@xt@\hsize{\hfil\box\@tempboxa\hfil}%
  \fi
  \vskip\belowcaptionskip}

\def\numberwithin#1#2{\@ifundefined{c@#1}{\@nocounterr{#1}}{%
  \@ifundefined{c@#2}{\@nocnterr{#2}}{%
  \@addtoreset{#1}{#2}%
  \toks@\@xp\@xp\@xp{\csname the#1\endcsname}%
  \@xp\xdef\csname the#1\endcsname
    {\@xp\@nx\csname the#2\endcsname
     .\the\toks@}}}}

{\theoremstyle{definition} \newtheorem{definition}{Definition} 
\newtheorem{example}{Example} 
 
\newtheorem{note}{Note}
}

\renewenvironment{thebibliography}[1]     
{\medskip
 \list{\@biblabel{\@arabic\c@enumiv}}%
           {\parsep=-0.2ex%
            \settowidth\labelwidth{\@biblabel{#1}}%
            \leftmargin\labelwidth
            \advance\leftmargin\labelsep
            \@openbib@code
            \usecounter{enumiv}%
            \let\p@enumiv\@empty
            \renewcommand\theenumiv{\@arabic\c@enumiv}}%
      \sloppy      \clubpenalty4000
      \@clubpenalty \clubpenalty
      \widowpenalty4000%
      \sfcode`\.\@m}
     {\def\@noitemerr
       {\@latex@warning{Empty `thebibliography' environment}}%
      \endlist}

\makeatother

\newcommand{\p}{\partial}

\newtheorem{myth}{Myth}
{\theoremstyle{definition} 
\newtheorem*{reformulation*}{Reformulation}
}

\begin{document}

\FirstPageHeading{kunzinger}

\ShortArticleName{Is a Nonclassical Symmetry a Symmetry?} 

\ArticleName{Is a Nonclassical Symmetry a Symmetry?}

\Author{Michael KUNZINGER~$^\dag$ and Roman O. POPOVYCH~$^\ddag$}
\AuthorNameForHeading{Michael Kunzinger and Roman O. Popovych}
\AuthorNameForContents{KUNZINGER M. and POPOVYCH R.O.}
\ArticleNameForContents{Is a nonclassical symmetry a symmetry?}

\Address{$^{\dag,\ddag}$Fakult\"at f\"ur Mathematik, Universit\"at Wien, Nordbergstra{\ss}e 15, A-1090\\
$\phantom{^\dag}$ Wien, Austria}
\EmailD{michael.kunzinger@univie.ac.at, rop@imath.kiev.ua}
\Address{$^\ddag$~Institute of Mathematics of NAS of Ukraine, 3 Tereshchenkivska Str.,\\
$\phantom{^\dag}$  Kyiv-4, Ukraine}


\Abstract{
Various versions of the definition of nonclassical symmetries existing in the literature are analyzed.
Comparing properties of Lie and nonclassical symmetries leads to the conclusion that in fact   
a nonclassical symmetry is not a symmetry in the usual sense. 
Hence the term ``reduction operator'' is suggested instead of the name ``operator of nonclassical symmetries''.
It is shown that in contrast to the case of single partial differential equations 
a satisfactory definition of nonclassical symmetries for systems of such equations has not 
been proposed up to now. 
Moreover, the cardinality of essential nonclassical symmetries is discussed, taking into account equivalence 
relations on the entire set of nonclassical symmetries. 
}

\vspace{-1.8ex}

\section{Introduction}

The ``nonclassical'' method of finding similarity solutions was introduced by Blu\-man and Cole in 1969 \cite{Bluman&Cole1969}. 
In fact, the method was first appeared in \cite{Bluman1967} in terms of ``nonclassical group'' 
but the terminology was changed in \cite{Bluman&Cole1969}.
Over the years the ``nonclassical'' method began to be associated with the term nonclassical symmetry \cite{Levi&Winternitz1989}
(also called $Q$-conditional \cite{Fushchych&Shtelen&Serov1989} or, simply, conditional symmetry \cite{Fushchych1987a,Fushchych&Zhdanov1992}).
In~the past two decades, the theoretical background of nonclassical symmetry was intensively investigated and  
nonclassical symmetry techniques were effectively applied to finding exact solutions 
of many partial differential equations arising in physics, biology, financial mathematics, etc. 
See, e.g., the review on investigations of nonclassical symmetries in~\cite{Olver&Vorob'ev1996}.

Here we mention only works which are directly connected with the subject of our paper. 
In the pioneering paper~\cite{Bluman&Cole1969} the ``nonclassical'' method was described 
by means of the example of the $(1+1)$-dimensional linear heat equation. 
It was emphasized that any solution of the corresponding (nonlinear) determining equations gives the coefficients of an operator 
such that an ansatz based on it reduces the heat equation to an ordinary differential equation. 
A veritable surge of interest in nonclassical symmetry was triggered by the papers \cite{Olver&Rosenau1986,Olver&Rosenau1987,Fushchych&Tsyfra1987}. 
In \cite{Olver&Rosenau1986} the ``nonclassical'' method was considered in the course of a comprehensive analysis of a wide range of 
methods for constructing exact solutions. 
The concept of weak symmetry of a system of partial differential equations, generalizing the ``nonclassical'' method, 
was introduced in~\cite{Olver&Rosenau1987}, 
where also the reduction procedure was discussed. 
Moreover, fundamental identities \cite[eq.~(23)]{Olver&Rosenau1987} crucially important for the theory of nonclassical symmetries 
were derived (see Myth~\ref{MythOnCriterionForNonclSym} below). 
The first version of the conditional invariance criterion explicitly taking into account differential consequences 
was proposed in~\cite{Fushchych&Tsyfra1987}. 
Generalizing results of~\cite{Fushchych&Tsyfra1987,Fushchych1987b} and other previous papers, 
in~\cite{Fushchych1987a} Fushchych introduced the notion of general conditional invariance. 
From the collection of papers containing~\cite{Fushchych1987a} it becomes apparent that 
around this time a number of authors began to regularly use the terms ``conditional invariance'' and ``$Q$-conditional invariance'' 
in connection with the method of Bluman and Cole.
The direct (ansatz) method closely related to this method was explicitly formulated in~\cite{Clarkson&Kruskal1989}. 
To the best of our knowledge, the name ``nonclassical symmetry'' was first used in~\cite{Levi&Winternitz1989}. 
Before this, there was no special name for operators calculated in this approach 
and the existing terminology on the subject emphasized characteristics of the method or invariance.  
The involution condition for families of operators
was first considered in the formulation of the conditional invariance criterion in \cite{Fushchych&Zhdanov1992,Vorob'ev1991}.  
The relations between nonclassical symmetries, reduction and formal compatibility of the combined system consisting of 
the initial equation and the invariant surface equation were discovered in~\cite{Pucci&Saccomandi1992} 
and were also studied in~\cite{Olver1994}.
The problem of the algorithmization of calculating nonclassical symmetries was posed in \cite{Clarkson&Mansfield1994a}. Furthermore, the
equivalence of the non-classical (conditional symmetry) and direct (ansatz) approaches to the reduction of partial differential equations 
was established in general form in \cite{Zhdanov&Tsyfra&Popovych1999},  
making use of the precise definition of reduction of differential equations.

In spite of the long history of nonclassical symmetry and the encouraging results in its applications, 
a number of basic problems of this theory are still open. 
Moreover, there exists a variety of non-rigorous definitions of related key notions 
and heuristic results on fundamental properties of nonclassical symmetry in the literature, 
which are used up to now and form what we would like to call the ``mythology'' of nonclassical symmetry.
These definitions and results require particular care and presuppose the tacit assumption of a number of
conventions in order to correctly apply them. 
Otherwise, certain contradictions and inaccurate statements may be obtained. 
Note that mythology interpreted in the above sense is an unavoidable and necessary step in the development of any subject.

Basic myths on nonclassical symmetries presented in the literature are discussed in this paper. 
We try to answer, in particular, the following questions.

\begin{itemize}
\item
Is a nonclassical symmetry a Lie symmetry of the united system 
of the initial equation and the corresponding invariant surface condition? 
Can a nonclassical symmetry be viewed as a conditional symmetry of the initial equation 
if the corresponding invariant surface condition is taken as the additional constraint? 
Is nonclassical symmetry a kind of symmetry in general? 
Does there exist a more appropriate name for this notion?
\item
What is a rigorous definition of nonclassical symmetry for systems of differential equations? 
Can such a definition be formulated as a straightforward extension of the definition of nonclassical symmetry 
for single partial differential equations? 
\item
Is the number of nonclassical symmetries essentially greater than the number of classical symmetries?  
\end{itemize}    

\section{Definition of nonclassical symmetry}

Following~\cite{Fushchych&Tsyfra1987,Fushchych&Zhdanov1992,Popovych&Vaneeva&Ivanova2007,Zhdanov&Tsyfra&Popovych1999}, 
in this section we briefly recall some basic notions and results on nonclassical (conditional) symmetries of 
partial differential equations. 
This will form the basis for our discussion of myths in the next sections.

Consider an involutive family $Q=\{Q^1,\ldots,Q^l\}$ of $l$ ($l\leqslant n$) first order differential operators 
(vector fields)
\[
Q^s=\xi^{si}(x,u)\p_i+\eta^s(x,u)\p_u, \quad s=1,\dots,l,
\]
in the space of the variables~$x$ and~$u$, satisfying the condition $\mathop{\rm rank}\nolimits \|\xi^{si}(x,u)\|=l$.

Here and in what follows
$x$ denotes the $n$-tuple of independent variables $(x_1,\ldots,$ $x_n)$, $n>1$, 
and $u$ is treated as the unknown function. 
The indices $i$ and $j$ run from 1 to $n$,  
the indices $s$ and $\sigma$ run from 1 to $l$, 
and we use the summation convention for repeated indices.
Subscripts of functions denote differentiation with respect to the corresponding variables, 
$\p_i=\p/\p x_i$ and $\p_u=\p/\p u$.
Any function is considered as its zero-order derivative.
All our considerations are in the local setting.

The requirement of involution for the family~$Q$ means that the commutator of any pair of operators from~$Q$ 
belongs to the span of $Q$ over the ring  of smooth functions of the variables~$x$ and $u$, i.e., 
\[
\forall\, s,s'\quad \exists \, \zeta^{ss'\sigma}=\zeta^{ss'\sigma}(x,u)\colon\quad [Q^s,Q^{s'}]=\zeta^{ss'\sigma}Q^\sigma.\]
The set of such families will be denoted by $\mathfrak Q^l$.

Consider an $r$th-order differential equation~$\mathcal L$ of the form~$L[u]:=L(x,u_{(r)})=0$
for the unknown function $u$ of the independent variables $x$.
Here, $u_{(r)}$ denotes the set of all derivatives of the function $u$ with respect to $x$
of order not greater than~$r$, including $u$ as the derivative of order zero.
Within the local approach the equation~$\mathcal L$ is treated as an algebraic equation 
in the jet space $J^r$ of the order $r$ and is identified with the manifold of its solutions in~$J^r$. 
Denote this manifold by the same symbol~$\mathcal L$ and 
the manifold defined by the set of all the differential consequences of the characteristic system~$Q[u]=0$ 
in $J^r$ by $\mathcal Q_{(r)}$, i.e., 
\begin{gather*}
\mathcal Q_{(r)}=\{ (x,u_{(r)}) \in J^r\, |\, D_1^{\alpha_1}\ldots D_n^{\alpha_n}Q^s[u]=0, 
\ \alpha_i\in\mathbb{N}\cup\{0\},\ |\alpha|<r \},
\end{gather*}
where 
$D_i=\p_{x_i}+u_{\alpha+\delta_i}\p_{u_\alpha}$ is the operator of total differentiation with respect to the variable~$x_i$, 
$Q^s[u]:=\eta^s-\xi^{si}u_i$ is the characteristic of the operator~$Q$, 
$\alpha=(\alpha_1,\ldots,\alpha_n)$ is an arbitrary multi-index, $|\alpha|:=\alpha_1+\cdots+\alpha_n$, 
$\delta_i$ is the multiindex whose $i$th entry equals 1 and whose other entries are zero. 
The variable $u_\alpha$ of the jet space $J^r$ corresponds to the derivative
$\p^{|\alpha|}u/\p x_1^{\alpha_1}\ldots\p x_n^{\alpha_n}$.

\begin{definition}\label{DefinitionOfNonclassicalSym}
The differential equation~$\mathcal L$ is called \emph{conditionally invariant} with respect to 
the involutive family $Q$ if the relation 
\begin{equation}\label{EqConditionalInvarianceCriterion}
Q^s_{(r)}L(x,u_{(r)})\bigl|_{\mathcal L\cap\mathcal Q_{\boldsymbol{(r)}}}=0
\end{equation}
holds,
which is called the \emph{conditional invariance criterion}.
Then $Q$ is called an \emph{involutive family of conditional symmetry} 
(or $Q$-conditional symmetry, nonclassical symmetry, etc.) operators of the equation~$\mathcal L$.
\end{definition}

Here the symbol $Q^s_{(r)}$ stands for the standard $r$th prolongation
of the operator~$Q^s$ \cite{Olver1993,Ovsiannikov1982}:
\[
Q^s_{(r)}=Q^s+\sum_{0<|\alpha|{}\leqslant  r} \bigl(D_1^{\alpha_1}\ldots D_n^{\alpha_n}Q^s[u]+\xi^{si}u_{\alpha+\delta_i}\bigr)\p_{u_\alpha}. 
\]

\section{Myths on name and definition}
\label{SectionOnMythsOnDefOfRedOps}

We restrict our consideration mainly to the case of families consisting of single operators ($l=1$) for simplicity and  
since mostly this case is investigated in the literature. 
Then the involution condition degenerates to an identity and we can omit the words ``involutive family'' and talk only about operators.  

\begin{myth}\label{MythOnNonclSymIsLieSym}
A nonclassical symmetry operator~$Q$ of an equation~$\mathcal L$ is a vector field~$Q$ 
which is a Lie symmetry operator of the united system of the equation~$\mathcal L$ 
and the invariant surface condition~$Q[u]=0$ corresponding to~$Q$. 
\end{myth}

This is the conventional non-rigorous way in order to quickly define nonclassical symmetry (see, e.g.,~\cite{Fushchych&Zhdanov1992,Hydon2000}). 
It becomes rigorous only after a special interpretation 
of the notions of system of differential equations and Lie symmetry. 
Otherwise, using the empiric definition leads to a number of inconsistencies.


A closer look reveals that the above definition is a tautology. 
Indeed, the invariant surface condition~$Q[u]=0$ means that the function $u$ is a fixed point of 
the one-parametric local group~$G_Q$ of local transformations generated by the operator~$Q$. 
Therefore, we can reformulate the definition in the following way. 

\begin{reformulation*}
If the set of those solutions of the equation~$\mathcal L$ which are fixed points of~$G_Q$, is invariant with respect to~$G_Q$, 
then $Q$ is called a nonclassical symmetry operator~$Q$ of the equation~$\mathcal L$. 
\end{reformulation*}

The tautology of the reformulation is obvious. 
If each element of the set is invariant then the whole set is necessarily invariant. 
The definition of nonclassical symmetry according to Myth~\ref{MythOnNonclSymIsLieSym} leads to the conclusion that 
\emph{any differential equation is invariant, in the nonclassical sense, with respect to 
any vector field in the corresponding space of dependent and independent variables.} 

The case when the equation~$\mathcal L$ has no $Q$-invariant solutions fits well into the non-rigorous approach 
in the sense that the empty set is a particularly symmetric~set. 

Therefore, uncritically following the non-rigorous approach, we would get 
no effective methods for constructing exact solutions and 
no information on the partial differential equations under consideration.

There exist a number reformulations of Myth~\ref{MythOnNonclSymIsLieSym} in the literature in different terms. 
The first one is in terms of conditional symmetry.

\begin{myth}\label{MythOnNonclSymIsCondSym}
A nonclassical symmetry operator~$Q$ of an equation~$\mathcal L$ is a conditional symmetry operator 
of the equation~$\mathcal L$ under the auxiliary condition~$Q[u]=0$. 
\end{myth}

The association of nonclassical symmetries (under the name $Q$-conditional symmetries) 
with conditional ones can be traced back to~\cite{Fushchych1987b} 
(see also~\cite{Fushchych&Shtelen&Serov1989} and earlier papers of the same authors). 
Here the term conditional symmetry is understood in the following sense~\cite{Fushchych1987a} 
(it can easily be defined for the case of a general system of differential equations). 

\begin{definition}\label{DefOfCondSym}
A vector field 
$Q$ is called a \emph{conditional symmetry operator} of a system~$\mathcal L$ of differential equations 
under an auxiliary condition~$\mathcal L'$ (which is another system of differential equations in the same variables) 
if $Q$ is a Lie symmetry operator of the united system of~$\mathcal L$ and~$\mathcal L'$.   
\end{definition}

Conditional symmetries defined in this way essentially differ from nonclassical symmetries. 
In particular, auxiliary conditions for conditional symmetries do not involve any associated conditional symmetry operators.
The conditional symmetry operators of a system~$\mathcal L$ under an auxiliary condition~$\mathcal L'$ form a Lie algebra.
Conditional symmetry indeed is a kind of symmetry and can be applied to generate new solutions from known ones. 
At the same time, in contrast to the case of nonclassical symmetries,
finding auxiliary conditions associated with nontrivial conditional symmetries is an art rather than an algorithmic procedure.
This is why sometimes nonclassical symmetries are called 
either \emph{$Q$-conditional symmetries}, where the prefix ``$Q$'' is used 
to emphasize the differences between nonclassical and conditional symmetries, 
or \emph{conditional symmetries} without any connection with Definition~\ref{DefOfCondSym}.

The second reformulation of Myth~\ref{MythOnNonclSymIsLieSym} is in infinitesimal terms. 
Note that infinitesimal criteria lie at the basis of Lie symmetry theory since 
they allow one to study linear problems for infinitesimal transformations 
instead of nonlinear problems for finite transformations. 

\begin{myth}\label{MythOnCriterionForNonclSym}
The conditional invariance criterion for an equation~$\mathcal L$ and an operator~$Q$ 
coincides with the infinitesimal Lie invariance criterion for the united system~$\{\mathcal L,Q[u]=0\}$
with respect to the same operator, i.e.,
\[
Q_{(r)}L[u]=0 \qquad\mbox{if}\qquad L[u]=0\quad \mbox{and}\quad Q[u]=0.
\]
\end{myth}
The infinitesimal Lie invariance criterion for the invariant surface condition $Q[u]=0$ 
with respect to the operator~$Q$ is identically satisfied as an algebraic consequence of this condition since
\[
Q_{(r)}Q[u]=Q_{(1)}Q[u]=(\eta_u-\xi^j_uu_j)Q[u]\equiv0 \quad\mbox{if}\quad Q[u]=0.
\]
We also have 
\begin{equation}\label{EqTautologicalRepresentationForProlongatedActionOfOpsOnEqs}
Q_{(r)}L[u]=\xi^iD_iL[u]+\sum_{|\alpha|{}\leqslant  r}L_{u_\alpha}[u]D_1^{\alpha_1}\ldots D_n^{\alpha_n} Q[u], 
\end{equation}
i.e., the equation $Q_{(r)}L[u]=0$ is a differential consequence of the equations \mbox{$L[u]=0$} and $Q[u]=0$ 
and, therefore, becomes an identity on the set of their common solutions. 
This tautology was first observed in~\cite{Olver&Rosenau1987}.

In the local approach to group analysis of differential equations, 
a system of differential equations is associated with the infinite tuple of 
systems of algebraic equations defined by this system and its differential consequences 
in the infinite tower of the corresponding jet spaces.  
The exclusion of the differential consequence~$Q_{(r)}L[u]$ when considering the system $L[u]=0$ and $Q[u]=0$
seems unnatural from the viewpoint of group analysis.

A variation of Myth~\ref{MythOnCriterionForNonclSym} is to replace, due to the Hadamard lemma,  
the ``invariance condition'' holding on the solution set of the system $L[u]=0$ and $Q[u]=0$ 
by the associated multiplier-condition, to be satisfied on the entire jet space $J^r$.

\begin{myth}\label{MythOnCriterionForNonclSymWithMultipliers}
An operator~$Q$ is a nonclassical symmetry of an equation~$\mathcal L$ if there exist $\lambda^1$ and $\lambda^2$ 
such that 
\begin{equation}\label{EqCondInvCriterionWithMultipliers}
Q_{(r)}L[u]=\lambda^1L[u]+\lambda^2Q[u].
\end{equation}
\end{myth}

The problem is to precisely define the nature of the multipliers $\lambda^1$ and $\lambda^2$. 
A~number of different conditions on the multipliers have been put forward in the literature. 
The simplest version is to prescribe no conditions at all on $\lambda^1$ and $\lambda^2$, 
which is obviously unacceptable. 

Sometimes $\lambda^1$ and $\lambda^2$ are assumed to be differential functions. 
This condition is natural for $\lambda^1$ but overly restrictive for $\lambda^2$. 
In fact, if only such $\lambda^2$ are allowed, 
the equivalence relation of nonclassical symmetries up to nonvanishing functional multipliers will be broken. 
Moreover, in this case the associated invariance criterion will become merely a sufficient condition for 
an ansatz constructed with the operator~$Q$ to reduce the equation~$\mathcal L$. 
As a result, a number of well-defined reductions may be lost.

On the other hand, requiring that both the multipliers $\lambda^1$ and $\lambda^2$ are 
polynomials of total differentiation operators 
with respect to the independent variables, whose coefficients are differential functions, 
is too weak an assumption. 
It arises from the association of nonclassical symmetries with conditional symmetries for which 
such multipliers are admissible. 
If we choose 
\[
\lambda^1=\xi^iD_i \quad\mbox{and}\quad \lambda^2=\sum_{|\alpha|{}\leqslant  r}L_{u_\alpha}[u]D_1^{\alpha_1}\ldots D_n^{\alpha_n},
\]
condition~\eqref{EqCondInvCriterionWithMultipliers} obviously becomes an identity for any operator~$Q$. 
In other words, condition~\eqref{EqCondInvCriterionWithMultipliers} reduces to 
the tautology~\eqref{EqTautologicalRepresentationForProlongatedActionOfOpsOnEqs} 
if both $\lambda^1$ and $\lambda^2$ are treated as differential operators of the above kind. 

Comparing Definition~\ref{DefinitionOfNonclassicalSym} and Myth~\ref{MythOnCriterionForNonclSymWithMultipliers} shows 
that $\lambda^1$ should be a differential function (i.e., a zeroth order operator) 
and $\lambda^2$ should be an order $(r-1)$ operator. 
These conditions for the multipliers can be weakened. 
Thus, bounding the order of total differentiations in~$\lambda^2$ is not essential. 
If $\lambda^1$ is a differential function,  condition~\eqref{EqCondInvCriterionWithMultipliers} implies 
that $\lambda^2$ cannot include total differentiations of orders greater than $r-1$. 
At the same time, explicitly prescribing the bound allows one to fix the order of the jet space under consideration.

\begin{myth}[The main myth of the theory]\label{MythOnNonclSymIsSym}
Nonclassical symmetry is a kind of symmetry of differential equations. 
\end{myth}

Any kind of symmetry of differential equations (Lie, contact, hidden, conditional, approximate, generalized, potential, nonlocal etc.)\ 
has the \emph{invariance} property, i.e., symmetries transform solutions to solutions in an appropriate sense.

The basic prerequisite of the definition of nonclassical symmetry is the consideration of only the set of solutions 
invariant under the associated finite transformations. 
It is impossible to use nonclassical symmetries in order to generate new solutions from known ones. 
A nonclassical symmetry operator~$Q$ of~$\mathcal L$ represents only a symmetry of 
\begin{itemize}\itemsep=0ex
\item
each $Q$-invariant solution of~$\mathcal L$ (as a weak symmetry~\cite{Olver&Rosenau1987}) and
\item
the manifold $\mathcal L\cap\mathcal Q_{(r)}$ in $J^{\boldsymbol{r}}$, 
where $r=\mathop{\rm ord}\mathcal L$.
\end{itemize}
The manifold $\mathcal L\cap\mathcal Q_{(r)}$ is properly related to the joint system $L[u]=0$ and $Q[u]=0$ 
of differential equations only if the operator~$Q$ and the equation~$\mathcal L$ satisfy the conditional invariance criterion.

At the same time, properties of the set of nonclassical symmetries 
and properties of the set of $Q$-invariant solutions for each nonclassical symmetry operator~$Q$ 
characterize the equation~$\mathcal L$.

Since a nonclassical symmetry is not in fact a kind of symmetry of differential equations, 
it is of utmost importance to discuss possibilities for replacing the name by one not involving the 
word ``symmetry''.

\section{Nonclassical symmetry, compatibility and reduction}

To understand the real nature of nonclassical symmetry, we discuss properties and applications of Lie symmetries and 
single out those of them which carry over to nonclassical symmetries.

\medskip

\noindent 
\textbf{Properties of Lie symmetries:}

\newcommand*\descriptionlabelMy[1]{\normalfont\it #1}

\begin{list}{}{
\labelwidth=2ex\labelsep=1ex\leftmargin=5mm
\topsep1mm\parsep1mm\itemsep0mm\partopsep0mm\let\makelabel\descriptionlabelMy}
\item[Invariance.] 
Any Lie symmetry (in the form of a parameterized family of finite transformations) locally maps the solution set 
of the corresponding system of differential equations onto itself. 
This is the main characteristic of any kind of symmetry. 
It gives rise to the possibility of generating new solutions from known ones. 
\item[Formal compatibility.] 
Attaching the invariant surface conditions associated with a Lie invariance algebra to the initial system of differential equations 
results in a system having no nontrivial differential consequences. 
In other words, the invariant surface conditions forms a class of proper universal differential constraints 
and, therefore, is appropriate for finding subsets of solutions of the initial system.
\item[Reduction.] 
Each Lie invariance algebra satisfying the infinitesimal transversality condition leads to 
an ansatz reducing the initial system to a system with a smaller number of independent variables, 
i.e., the reduced system is more easily solvable than the initial one.   
\item[Conditional compatibility.] 
There exists a bijection between solutions of the initial system which satisfy the invariant surface conditions, 
and solutions of the corresponding reduced system. 
This means that all solutions of the initial system invariant with respect to a Lie invariance algebra, 
can be constructed via solving the corresponding reduced system. 
\end{list}

For nonclassical symmetries, the property of invariance is broken but the other properties 
(formal compatibility, reduction, conditional compatibility) are preserved. 
In fact, the conditional invariance criterion~\eqref{EqConditionalInvarianceCriterion} is 
the condition of formal compatibility of the joint system $L[u]=0$ and $Q[u]=0$~\cite{Pucci&Saccomandi1992}.  
We can identify \emph{nonclassical symmetries of~$\mathcal L$} 
with \emph{first-order quasilinear differential constraints which are formally compatible with $\mathcal L$}.%
\footnote{%
In fact, this claim and Definition~\ref{DefinitionOfNonclassicalSym2} are not entirely rigorous. 
Their precise formulation depends on what definition of formal compatibility is used. 
Consider, e.g., the definition presented in \cite{Seiler1994,Seiler2010}. 
We temporarily use notations compatible with these references, hence slightly different from the rest of the paper.  

Let $\mathcal L_r$ be a system of $l$~differential equations $\smash{L^1[u]=0}$, \dots, $\smash{L^l[u]=0}$ 
in $n$~independent variables $\smash{x=(x_1,\dots,x_n)}$ and
$m$~dependent variables $\smash{u=(u^1,\dots,u^m)}$, which involves derivatives of~$u$ up to order~$r$. 
The system~$\mathcal L_r$ is interpreted as a system of algebraic equations in the jet space~$J^r$ 
and defines a manifold in~$J^r$, which is also denoted by~$\mathcal L_r$. 
The $s$th order prolongation $\mathcal L_{r+s}$ of the system~$\mathcal L_r$, $s\in\mathbb N$, is the system in~$J^{r+s}$ 
consisting of the equations  $\smash{D_1^{\alpha_1}\ldots D_n^{\alpha_n}L^k[u]=0}$, $k=1,\dots,l$, $|\alpha|\leqslant s$.
The projection of the corresponding manifold on $J^{r+s-q}$, where $q\in\mathbb N$ and $q\leqslant s$, is denoted by $\smash{\mathcal L_{r+s-q}^{(q)}}$. 
The system~$\mathcal L_r$ is called \emph{formally integrable} (or \emph{formally compatible}) if 
$\smash{\mathcal L_{r+s}^{(1)}=\mathcal L_{r+s}}$ for any $s\in\mathbb N$ \cite{Seiler1994,Seiler2010}. 

The first obstacle in the harmonization of the above definition of formal compatibility and the definition of nonclassical symmetry 
is that the equations $L[u]=0$ and $Q[u]=0$ have, as a rule, different orders. 
Therefore, differential consequences of the equation $Q[u]=0$ should be attached to the joint system $L[u]=0$ and $Q[u]=0$ 
before testing its compatibility. 
The second obstacle is that the order of $L[u]$ may be lowered on the manifold $\smash{\mathcal Q_{(r)}}$ 
if $Q$ is a singular vector field for the equation $L[u]=0$. 
Hence instead of the equation $L[u]=0$ we should use the equation $L_*[u]=0$, where 
$L_*$ is a differential function 
which coincides with~$\lambda L$ on $\smash{\mathcal Q_{(r)}}$ 
for some nonvanishing differential function~$\lambda$ 
and whose order~$r_*$ is minimal among differential functions possessing this property. 
Finally, we arrive at the following definition:
The differential equation~$\mathcal L$ is called \emph{conditionally invariant} with respect to 
the involutive family of operators $Q$ 
if the system $L_*[u]=0$, $\smash{D_1^{\alpha_1}\ldots D_n^{\alpha_n}Q[u]=0}$, $|\alpha|<r_*$, is formally compatible.
}

\begin{definition}\label{DefinitionOfNonclassicalSym2}
The differential equation~$\mathcal L$ is called \emph{conditionally invariant} with respect to 
the involutive family of operators $Q$ if the joint system of~$\mathcal L$ with the characteristic system~$Q[u]=0$ is formally compatible.
\end{definition}

What is the main property that adequately represents the essence of nonclassical symmetry?

The fact that the characteristic equations $Q^s[u]=0$ are quasilinear and of first order implies
the possibility of integrating them explicitly, 
i.e., an ansatz associated with the characteristic system~$Q[u]=0$ can be constructed. 
In view of the Frobenius theorem, the involution and transversality conditions for the family~$Q$ 
(together with the fact that the operators from~$Q$ are of first order) 
imply that the ansatz involves one new unknown function of $n-l$ new independent variables. 
Then the formal compatibility of the joint system $L[u]=0$ and $Q[u]=0$ guaranties 
the reduction of~$\mathcal L$ by the ansatz to a single differential equations $\mathcal L'$ in $n-l$ independent variables. 
Thus, the number of dependent variables and equations are preserved under the reduction with~$Q$
and the number of independent variables decreases by the cardinality of~$Q$, i.e.,  
similarly to Lie symmetries nonclassical symmetries lead to
the conventional reduction of the number of independent variables.  

There exist integrable differential constraints which are not formally compatible with the initial system.
Differential constraints can be formally compatible with the initial system and, at the same time, non-integrable in an explicit form. 
An ansatz constructed with a general integrable differential constraint may involve a number of new unknown functions 
depending on different variables. 
Therefore, only all the above properties combined (first order, quasilinearity, formal compatibility,  transversality and involution) 
result in the classical reduction procedure.%
\footnote{Extended notions of reduction are also used. 
Thus, weak symmetries imply reductions decreasing the number of independent variables, 
preserving the number of unknown functions and increasing the number of equations~\cite{Olver&Rosenau1987}. 
The reduced system can be much more overdetermined than the initial one. 
The reductions associated with higher-order nonclassical symmetries preserve the determinacy type of systems, 
simultaneously increasing the numbers of unknown functions and  equations~\cite{Olver1994}.} 

The conditional invariance of the equation~$\mathcal L$ with respect to the family~$Q$
is equivalent to the ansatz constructed with this family reducing~$\mathcal L$
to a differential equation with $n-l$ independent variables~\cite{Zhdanov&Tsyfra&Popovych1999}. 
Moreover, reducing the number of independent variables in partial differential equations 
is the main goal in the study of nonclassical symmetries.  
Since the reduction by the associated ansatz is the quintessence of nonclassical symmetries, 
it was proposed in~\cite{Popovych2008a,Popovych&Vaneeva&Ivanova2007,Vasilenko&Popovych1999} 
to call involutive families of nonclassical symmetry operators  
{\it families of reduction operators} of~$\mathcal L$.
 
Another important property holding for Lie symmetries is broken for nonclassical symmetries.
Let the equation~$\mathcal  L$ be of order~$r$ and \[L_{(k)}=\{ D_1^{\alpha_1}\ldots D_n^{\alpha_n}L[u]=0,\ |\alpha|\leqslant k-r\}.\]  
Denote by $L_{(k)}$ a maximal set of algebraically independent differential consequences of~$\mathcal L$
that have, as differential equations, orders not greater than $k$. 
We identify $L_{(k)}$ with the corresponding system of algebraic equations in~$J^k(x|u)$ 
and associate it with the manifold $\mathcal L_{(k)}$ determined by this system. 
For Lie symmetries we have the following properties.
\begin{list}{\arabic{enumi}.}{\usecounter{enumi}
\labelwidth=2ex\labelsep=1ex\leftmargin=3ex
\topsep1mm\parsep1mm\itemsep0mm\partopsep0mm}
\item
If $Q$ is a Lie symmetry operator of $\mathcal L_{(r)}$ then $Q$ is a Lie symmetry operator of $\mathcal L_{(\rho)}$ for any $\rho>r$.
\item
If $Q$ is a Lie symmetry operator of $\mathcal L_{(\rho)}$ for some $\rho>r$ then $Q$ is a Lie symmetry of $\mathcal L_{(r)}$.
\end{list}

The first of these properties extends to nonclassical symmetries but this is not the case for the second one. 
In fact: 
\begin{list}{\arabic{enumi}.}{\usecounter{enumi}
\labelwidth=2ex\labelsep=1ex\leftmargin=3ex
\topsep1mm\parsep1mm\itemsep0mm\partopsep0mm}
\item
If $Q$ is a Lie symmetry operator of $\mathcal L_{(r)}\cap\mathcal Q_{(r)}$ 
then $Q$ is a Lie symmetry operator of $\mathcal L_{(\rho)}\cap\mathcal Q_{(\rho)}$ for any $\rho>r$.
\item
The fact that $Q$ is a Lie symmetry operator of $\mathcal L_{(\rho)}\cap\mathcal Q_{(\rho)}$ for some $\rho>r$ 
does not imply that $\mathcal L_{(r)}\cap\mathcal Q_{(r)}$ admits the operator~$Q$.
\end{list}

\begin{example}\label{ExampleOnNonclassicalSymsOfProlongation} 
Let $L[u]=u_t+u_{xx}+tu_x$, $\mathcal L\colon L[u]=0$ and $Q=\p_t$. 
Then the manifold $\mathcal L_{(2)}\cap\mathcal Q_{(2)}$ is determined in $J^2$ by the equations $u_t=u_{tt}=u_{tx}=0$ and $u_{xx}=-tu_x$.
Since \[Q_{(2)}L\bigl|_{\mathcal L_{(2)}\cap\mathcal Q_{(2)}}=u_x\not=0,\] 
the operator $\p_t$ is not a reduction operator of $\mathcal L$. 
Substituting the corresponding ansatz $u=\varphi(\omega)$, where the invariant independent variable is $\omega=x$, into~$\mathcal L$ 
results in the equation~$\varphi_{\omega\omega}+t\varphi_\omega=0$, in which the ``parametric'' variable~$t$ cannot be excluded 
via multiplying by a nonvanishing differential function. 
As expected, the ansatz does not reduce the equation~$\mathcal L$.

Consider the same operator~$Q$ and the first prolongation~$\mathcal L_{(3)}$ of~$\mathcal L$, 
which is determined by the equations $L[u]=0$,  $D_tL[u]=0$ and $D_xL[u]=0$. 
The manifold $\mathcal L_{(3)}\cap\mathcal Q_{(3)}$ is singled out from~$J^3$ by the equations
\[
u_t=u_{tt}=u_{tx}=u_{ttt}=u_{ttx}=u_{txx}=0,\quad u_x=u_{xx}=u_{xxx}=0.
\]
The conditional invariance criterion is satisfied for the prolonged system~$\mathcal L_{(3)}$ and the operator~$Q$: 
\[
Q_{(2)}L\bigl|_{\mathcal L_{(3)}\cap\mathcal Q_{(3)}}=
Q_{(3)}D_tL\bigl|_{\mathcal L_{(3)}\cap\mathcal Q_{(3)}}=
Q_{(3)}D_xL\bigl|_{\mathcal L_{(3)}\cap\mathcal Q_{(3)}}=0, 
\]
i.e., $Q$ is a nonclassical symmetry operator of the system~$\mathcal L_{(3)}$ and the above ansatz reduces~$\mathcal L_{(3)}$ to 
the system of three ordinary differential equations $\varphi_\omega=0$, $\varphi_{\omega\omega}=0$ and $\varphi_{\omega\omega\omega}=0$ since 
\begin{gather*}
\left(\begin{array}{c}\varphi_{\omega\omega}+t\varphi_\omega\\\varphi_\omega\\\varphi_{\omega\omega\omega}+t\varphi_{\omega\omega}\end{array}\right)=
\left(\begin{array}{ccc}t&1&0\\1&0&0\\0&t&1\end{array}\right)\!
\left(\begin{array}{c}\varphi_{\omega\omega}\\\varphi_{\omega\omega}\\\varphi_{\omega\omega\omega}\end{array}\right)=0 
\quad\!\mbox{and}\quad\! \left|\begin{array}{ccc}t&1&0\\1&0&0\\0&t&1\end{array}\right|\ne0.
\end{gather*}
\end{example}

\begin{note}\label{NoteOnNonclassicalSymsOfProlongation} 
In general, for any system~$\mathcal L$ and any involutive family~$Q$ there exists an order~$r$ such that $\mathcal L_{(r)}\cap\mathcal Q_{(r)}$ 
is invariant with respect to~$Q_{(r)}$. 
This gives the theoretical background of the notion of weak symmetry \cite{Olver&Rosenau1987}.   
\end{note}

\section{Definition of nonclassical symmetries for systems}\label{SectionOnMythsOnSystems}

\begin{myth}\label{MythOnDefOfNoclSymForSystems}
The definition of nonclassical symmetry for systems of differential equations is 
a simple extension of the definition of nonclassical symmetry 
for single partial differential equations to the case of systems. 
\end{myth}

Example~\ref{ExampleOnNonclassicalSymsOfProlongation} and Note~\ref{NoteOnNonclassicalSymsOfProlongation} 
indicate problems arising in attempts of defining nonclassical symmetries for systems of partial differential 
equations. 

Let~$\mathcal L$ denote a system~$L(x,u_{(r)})=0$ of $l$ differential equations $L^1=0$, \ldots, $L^l=0$
for $m$ unknown functions $u=(u^1,\ldots,u^m)$
of $n$ independent variables $x=(x_1,\ldots,x_n)$.
It is always assumed that the set of differential equations forming the system under consideration 
canonically represents this system and is minimal. 
The minimality of a set of equations means that no equation from this set is a differential consequence of the other equations.
By $L_{(k)}$ we will denote a maximal set of algebraically independent differential consequences of~$\mathcal L$
that have, as differential equations, orders not greater than $k$. 
We identify~$L_{(k)}$ with the corresponding system of algebraic equations in the jet space~$J^k$ and 
associate it with the manifold~$\mathcal L_{(k)}$ determined by this system. 
Let $L_{(r)}=\{\hat L^\nu, \nu=1,\dots,\hat l\}$.
Note that the general system includes equations of different orders. 

What is the correct conditional invariance criterion for the system~$\mathcal L$?
\begin{gather*}
Q_{(r)}L^\mu\bigl|_{\mathcal L\cap\mathcal Q_{(r)}}=0,\quad \mu=1,\dots,l\,?
\\ 
Q_{(r)}L^\mu\bigl|_{\mathcal L_{(r)}\cap\mathcal Q_{(r)}}=0,\quad \mu=1,\dots,l\,? 
\\ 
Q_{(r)}\hat L^\nu\bigl|_{\mathcal L_{(r)}\cap\mathcal Q_{(r)}}=0,\quad \nu=1,\dots,\hat l\,? 
\end{gather*}

All of the above candidates for the criterion are not satisfactory. 
The second candidate is not a good choice 
since it neglects the equations having lower orders than the order of the whole system. 
Taking the third candidate, we obtain nonclassical symmetries of a prolongation of the system. 
As shown by Example~\ref{ExampleOnNonclassicalSymsOfProlongation}, 
these may be weakly related to nonclassical symmetries of the system. 
It is not well understood what differential consequences are really essential. 
Thus, elements of~$\mathcal L_{(r)}$ whose trivial differential consequences also belong to~$\mathcal L_{(r)}$ 
are neglected by this candidate.

Although all operators satisfying the first of the above criteria give proper reductions, 
it is overly restrictive and in fact is only a sufficient condition for nonclassical symmetries. 
Even Lie symmetries can be lost when employing it. 

The above discussion is illustrated by the following example. 

\begin{example}
Consider the system
\begin{equation}\label{EqModifiedNSEs}
\vec u_t+(\vec u\cdot\nabla)\vec u-\Delta\vec u+\nabla p+\vec x\times\nabla\mathop{\rm div}\vec u=\vec 0, \quad \mathop{\rm div}\vec u=0.
\end{equation}
which is obviously equivalent to the system of Navier--Stokes equations describing the motion of an incompressible fluid. 
(The additional term $\vec x\times\nabla(\mathop{\rm div}\vec u)$ vanishes if $\mathop{\rm div}\vec u=0$.) 
If we do not take into account differential consequences of system~\eqref{EqModifiedNSEs}, 
we derive the unnatural claim that this system is not conditionally invariant with respect to translations of the space variables~$x_i$.
At the same time, the infinitesimal generators of these translations belong to 
the maximal Lie invariance algebra of the Navier--Stokes equations. 
A maximal set $L_{(2)}$ of algebraically independent differential consequences of~$\mathcal L$
that have, as differential equations, orders not greater than $2$ 
is formed by the equations 
\begin{gather*}
\vec u_t+(\vec u\cdot\nabla)\vec u-\Delta\vec u+\nabla p=\vec 0, \quad \mathop{\rm div}\vec u=0, 
\\[1ex]
\mathop{\rm div}\vec u_t=\vec 0, \quad \nabla\mathop{\rm div}\vec u=\vec 0, \quad u^i_ju^j_i+\Delta p=0.
\end{gather*}
Here the indices $i$ and~$j$ run from~1 to~3. 
The equation $Q_{(2)}\mathop{\rm div}\vec u=\vec 0$ is identically satisfied on the set 
$\mathcal L_{(2)}\cap\mathcal Q_{(2)}$. 
Therefore, the application of the second or third candidate for the conditional invariance criterion to the equation 
$\mathop{\rm div}\vec u=\vec 0$ gives no determining equations for nonclassical symmetries of the system~\eqref{EqModifiedNSEs}.
\end{example}

Definition~\ref{DefinitionOfNonclassicalSym2} can also not be extended to the case of systems in an easy way. 
The problem again is to define what set of differential consequences of the initial system should be chosen 
for testing formal compatibility with the appropriate characteristic system. 

The notions of nonclassical symmetry and reduction are strongly related in the case of single partial differential equations.
It therefore seems natural for these notions to also be closely related in the case of systems. 
Hence the problem of rigorously defining nonclassical symmetries for systems is additionally complicated 
by the absence of a canonical extension of the classical reduction to the case of systems.
A chain of simple examples can be presented to illustrate possible features of such an extension.

\section{Myths on number of nonclassical symmetries}
\label{SectionOnMythsOnSingularCasesAndFactorizationAndNumber}

\begin{myth}\label{MythOnNumber}
The number of nonclassical symmetries is essentially greater than the number of classical symmetries.  
\end{myth}

At first sight this statement seems obviously true. 
There exist classes of partial differential equations 
whose maximal Lie invariance algebra is zero and which admit large sets of reduction operators.
This is the case, e.g., for general $(1+1)$-dimensional evolution equations. 
At the same time, certain circumstances significantly reduce the number of essential nonclassical symmetries. 
We briefly list them below. 

\begin{itemize}\itemsep=-.5ex
\item
The usual equivalence of families of reduction operators. 
Involutive families~$Q$ and $\widetilde Q$ of $l$ operators are called equivalent if 
$\widetilde Q^s=\lambda^{s\sigma}Q^\sigma$ for some $\lambda^{s\sigma}=\lambda^{s\sigma}(x,u)$ with $\det\|\lambda^{s\sigma}\|\not=0$.
\item
Nonclassical symmetries equivalent to Lie symmetries.
\item 
The equivalence of nonclassical symmetries with respect to Lie symmetry groups of single differential equations \cite{Levi&Winternitz1989,Popovych2000}
and equivalence groups of classes of such equations~\cite{Popovych&Vaneeva&Ivanova2007}. 
\item 
No-go cases. The problem of finding certain wide subsets of reduction operators may turn out to be 
equivalent to solving the initial equation 
\cite{Kunzinger&Popovych2008,Popovych2008a}.
\item 
Non-Lie reductions leading to Lie invariant solutions. 
\end{itemize}

Thus, the existence of a wide Lie symmetry group for a partial differential equation~$\mathcal L$ complicates, in a certain sense,  
finding nonclassical symmetries of~$\mathcal L$. 
Indeed, any subalgebra of the corresponding maximal Lie invariance algebra, satisfying the transversality condition, 
generates a class of equivalent Lie families of reduction operators. 
If a non-Lie family of reduction operators exists, the action of symmetry transformations on it results in  
a series of non-Lie families of reduction operators, which are inequivalent in the usual sense. 
Therefore, for any fixed value of~$l$ the system of determining equations for the coefficients of operators 
from the set $\mathfrak Q^l(\mathcal L)$ of families of $l$~reduction operators 
is not sufficiently overdetermined to be completely integrated in an easy way, 
even after factorizing with respect to the equivalence relation in $\mathfrak Q^l(\mathcal L)$. 
To produce essentially different non-Lie reductions, one has to exclude the solutions of the determining equations  
which give Lie families of reduction operators and non-Lie families which are equivalent to others with respect to 
the Lie symmetry group of~$\mathcal L$. 
As a result, the ratio of efficiency of such reductions to the expended efforts can become vanishingly small.

\section{Conclusion}

Although the name ``nonclassical symmetry'' and other analogous names for reduction operators, 
which refer to symmetry or invariance, do not reflect actual properties of these objects, 
the usage of such names is justified by historical conventions and additionally
supported by the terminology of related fields of group analysis of differential equations. 
It is a quite common situation for different fields of human activity that 
a modifier completely changes the meaning of the initial notion
(think of terms like ``negative growth'', ``military intelligence'', etc.).
Empiric definitions of nonclassical symmetry can be used in a consistent way
if all involved terms and notions are properly interpreted. 
Nevertheless, as we have argued, the term reduction operator more adequately
captures the underlying mathematical content. 

In this paper we discussed certain basic myths of the theory of nonclassical symmetries, 
pertaining to different versions of their definition and the estimation of their cardinality. 
Over and above these, there are a number of more sophisticated myths concerning, among others, the factorization of sets of nonclassical symmetries, 
involutive families of reduction operators in the multidimensional case, and
singular sets of reduction operators. 
A discussion of such myths requires a careful theoretical analysis substantiated by nontrivial examples and 
will be the subject of a forthcoming paper.

\subsection*{Acknowledgements}

ROP thanks the members of the local organizing committee for the nice conference and for their hospitality.
MK was supported by START-project Y237 of the Austrian Science Fund. 
The research of ROP was supported by the Austrian Science Fund (FWF), project P20632.

\LastPageEnding


\begin{thebibliography}{99}
\footnotesize

\bibitem{ArrigoHillBroadbridge1993}
Arrigo D.J., Hill J.M., Broadbridge P.,
Nonclassical symmetry reductions of the linear diffusion equation with a nonlinear source,
{\it IMA J. Appl. Math.}, 1994, {\bf 52}, 1--24.

\bibitem{Bluman1967}
Bluman~G.W. Construction of solutions to partial differential equations by the use of transformation groups, Thesis, 
California Institute of Technology, 1967.

\bibitem{Bluman&Cole1969}
Bluman~G.W. and Cole~J.D.,
The general similarity solution of the heat equation,
{\it J. Math. Mech.}, 1969, {\bf 18}, 1025--1042.

\bibitem{Clarkson&Kruskal1989}
Clarkson P.A. and Kruskal M.D., 
New similarity solutions of the Boussinesq equation, 
{\it J. Math. Phys.}, 1989, {\bf 30}, 2201--2213.

\bibitem{Clarkson&Mansfield1994a}
Clarkson P.A. and Mansfield E.L., 
Algorithms for the nonclassical method of symmetry reductions, 
{\it SIAM J. Appl. Math.},  1994, {\bf 54}, 1693--1719. 

\bibitem{Fushchych1987a}
Fushchych W.I., 
How to extend symmetry of differential equations? 
in  \emph{Symmetry and Solutions of Nonlinear Equations of Mathematical Physics}, Kyiv, Inst. of Math. Acad. of Sci. of Ukraine, 1987, 4--16. 

\bibitem{Fushchych1987b}
Fushchych W.I., 
On symmetry and exact solutions of multi-dimensional nonlinear wave equations, 
Ukrain. Math. J., 1987, V.39, N 1, p.116Ц123 (in Russian). 

\bibitem{Fushchych&Shtelen&Serov1989}
Fushchych W.I., Shtelen W.M. and Serov N.I., 
{\it Simmetrijnyj analiz i tochnye resheniya nelinejnykh uravnenij matematicheskoj fiziki}, Kyiv, Naukova Dumka, 1989 (in Russian). 
Translated and extended version:
{\it Symmetry analisys and exact solutions of equations of nonlinear mathimatical physics},  
Dordrecht, Kluwer Academic Publishers, 1993. 

\bibitem{Fushchych&Tsyfra1987}
Fushchych W.I. and Tsyfra I.M.,
On a reduction and solutions of the nonlinear wave equations with broken symmetry,
{\it J. Phys. A: Math. Gen.}, 1987, {\bf 20}, L45--L48.

\bibitem{Fushchych&Zhdanov1992} 
Fushchych~W.I. and Zhdanov~R.Z., 
Conditional symmetry and reduction of partial differential equations, 
{\it Ukr. Math. J.}, 1992, {\bf 44}, 970--982.

\bibitem{Hydon2000}
Hydon P., 
{\it Symmetry Methods for Differential Equations: A Beginner's Guide}, 
Cambridge University Press, 2000.

\bibitem{Kunzinger&Popovych2008}
Kunzinger M. and Popovych R.O., 
Singular reduction operators in two dimensions, 
{\it J. Phys. A}, 2008, {\bf 41}, 505201, 24 pp, arXiv:0808.3577. 

\bibitem{Levi&Winternitz1989}
Levi D. and Winternitz P., 
Non-classical symmetry reduction: example of the Boussinesq equation
{\it J. Phys. A: Math. Gen.}, 1989, {\bf 22}, 2915--2924.

\bibitem{Olver1993}
Olver P., {\it Applications of Lie groups to differential equations},
New-York, Springer-Verlag, 1993.

\bibitem{Olver1994}
Olver P., Direct reduction and differential constraints,
{\it Proc. R. Soc. Lond. A}, 1994, {\bf 444}, 509--523.

\bibitem{Olver&Rosenau1986}
Olver P.J. and Rosenau P., 
The construction of special solutions to partial differential equations,
{\it Phys. Lett. A}, 1986, {\bf 114}, 107--112.

\bibitem{Olver&Rosenau1987}
Olver P.J. and Rosenau P., 
Group-invariant solutions of differential equations,
{\it SIAM J. Appl. Math.}, 1987, {\bf 47}, 263--278.

\bibitem{Olver&Vorob'ev1996}
Olver P.J. and Vorob'ev E.M., Nonclassical and conditional symmetries, 
in {\it CRC Handbook of Lie Group Analysis of Differential Equations}, 
Vol. 3, Editor N.H. Ibragimov, Boca Raton, Florida, CRC Press, 1996, 291--328.

\bibitem{Ovsiannikov1982}
Ovsiannikov~L.V., {\it Group analysis of differential equations},
New York, Academic Press, 1982.

\bibitem{Popovych2000}
Popovych R.O., 
Equivalence of $Q$-conditional symmetries under group of local transformation,
in Proceedings of the Third International Conference ``Symmetry in Nonlinear Mathematical Physics'' (Kyiv, July 12-18, 1999),
{\it Proceedings of Institute of Mathematics}, Kyiv, 2000, {\bf 30}, Part 1, 184--189; arXiv:math-ph/0208005.

\bibitem{Popovych2008a}
Popovych R.O., Reduction operators of linear second-order parabolic equations,  
{\it J. Phys. A}, 2008, {\bf 41}, 185202; arXiv:0712.2764.

\bibitem{Popovych&Vaneeva&Ivanova2007}
Popovych R.O., Vaneeva O.O and Ivanova N.M., 
Potential nonclassical symmetries and solutions of fast diffusion equation, 
{\it Phys. Lett. A}, 2007, {\bf 362}, 166--173; arXiv:math-ph/0506067.

\bibitem{Pucci&Saccomandi1992} 
Pucci E. and Saccomandi G.,
On the weak symmetry groups of partial differential equations, 
{\it J. Math. Anal. Appl.}, 1992, {\bf 163}, 588--598.

\bibitem{Seiler1994} 
Seiler W.M., 
{\it Analysis and application of the formal theory of partial differential equations}, 
thesis, Lancaster University, 1994.

\bibitem{Seiler2010} 
Seiler W.M., 
{\it Involution. The Formal Theory of Differential Equations and its Applications in Computer Algebra}, 
Springer-Verlag, Berlin Heidelberg, 2010.

\bibitem{Vasilenko&Popovych1999}
Vasilenko O.F. and Popovych R.O., 
On class of reducing operators and solutions of evolution equations, 
{\it Vestnik PGTU}, 1999, {\bf 8}, 269--273 (in Russian).

\bibitem{Vorob'ev1991}
Vorob'ev E.M., 
Reduction and quotient equations for differential equations with symmetries, 
{\it Acta Appl. Math.}, 1991, {\bf 51}, 1--24.

\bibitem{Zhdanov&Tsyfra&Popovych1999}
Zhdanov R.Z., Tsyfra I.M. and Popovych R.O.,
A precise definition of reduction of partial differential equations,
{\it J. Math. Anal. Appl.}, 1999, {\bf 238}, 101--123; arXiv:math-ph/0207023.

\end{thebibliography}
\end{document}